\documentclass[fleqn,usenatbib]{mnras}

\usepackage{newtxtext,newtxmath}
\usepackage{hyperref}

\usepackage[T1]{fontenc}

\DeclareRobustCommand{\VAN}[3]{#2}
\let\VANthebibliography\thebibliography
\def\thebibliography{\DeclareRobustCommand{\VAN}[3]{##3}\VANthebibliography}

\usepackage{graphicx}	
\usepackage{amsmath}	

\newcommand{\mm}[1]{\mbox{$#1$}} 
\newcommand{\betaf}{\mm{\beta/f}}
\newcommand{\Msol}{\mm{M_\odot}}
\newcommand{\h}{\mm{h^{-1}}}
\newcommand{\fsig}{\mm{f\sigma_8}}
\newcommand{\sig}{\mm{\sigma_8}}
\newcommand{\bstar}{\mm{\beta^*}}
\newcommand{\bm}[1]{\boldsymbol{#1}}
\newcommand{\sbr}[1]{_{\textrm{#1}}} 

\title[Testing Accuracy of Peculiar Velocity Comparisons]%
{Assessing the accuracy of cosmological parameters estimated from velocity -- density comparisons via simulations}

\author[Hollinger and Hudson]{
Amber M. Hollinger$^{1,2}$
 and Michael J. Hudson$^{1,2,3}$
\\
$^{1}$Department of Physics and Astronomy, University of Waterloo, 200 University Ave W, Waterloo, ON N2L 3G1, Canada\\
$^{2}$Waterloo Centre for Astrophysics, University of Waterloo, 200 University Ave W, Waterloo, ON N2L 3G1, Canada\\
$^{3}$Perimeter Institute for Theoretical Physics, 31 Caroline St. North, Waterloo, ON N2L 2Y5, Canada
}

\date{Accepted XXX. Received YYY; in original form ZZZ}

\pubyear{2020}

\begin{document}
\label{firstpage}
\pagerange{\pageref{firstpage}--\pageref{lastpage}}
\maketitle

\begin{abstract}

A promising method for measuring the cosmological parameter combination $f\sigma_8$ is to compare observed peculiar velocities with peculiar velocities predicted from a galaxy density field using perturbation theory.
We use $N$-body simulations and semi-analytic galaxy formation models to quantify the accuracy and precision of this method. Specifically, we examine a number of technical aspects, including the optimal smoothing length applied to the density field, the use of dark matter halos or galaxies as tracers of the density field, the effect of noise in the halo mass estimates or in the stellar-to-halo mass relation, and the effect of finite survey volumes. We find that for a Gaussian smoothing of $4 \h$ Mpc, the method has only small systematic biases at the level of 5\%. Cosmic variance affects current measurements at the 5\% level due to the volume of current redshift data sets.

\end{abstract}




\section{Introduction}

Peculiar velocities are the only practical way of measuring the underlying distribution of dark matter (DM) on large scales \citep{willick1997homogeneous}, in the nearby (low redshift) Universe. Specifically, they can be used to measure the cosmological parameter combination $f(\Omega_m) \sigma_8$, where the first term is the logarithmic growth rate of fluctuations, with $f(\Omega_m) = \Omega_m^{0.55}$ in the Lambda Cold Dark Matter ($\Lambda$CDM) model, and $\sigma_8$ is the root mean square matter density fluctuation in a sphere of radius 8 \h\ Mpc. Although peculiar velocities have been used for similar purposes since the early 90s (see reviews by \cite{Dek94} and \cite{StrWil95}), there has been a recent revival of interest in measuring \fsig\ \citep{pike2005cosmological,DavNusMas11,turnbull_sn, hudson2012growth, carrick2015cosmological,Huterer17,cosmicflows_growth_rate, 2019MNRAS.487.5235Q, 6df_cross_w_rsd,SaiColMag20,BorHudLav20}.

The use of peculiar velocities as a probe of \fsig\ has taken on renewed importance in light of the 3.2$\sigma$ conflict between Cosmic Microwave Background \citep{planck_cmb_cosmo} and weak lensing measurements \citep{asgari2019kidsviking450} of the parameter combination $S_8 \equiv \Omega_m^{0.5}\sigma_8$, with the latter giving lower values. This combination is very similar to \fsig, differing by only 0.05 in the exponent of $\Omega_m$. As shown by \cite{BorHudLav20}, some recent peculiar velocity measurements also yield lower values of $S_8$ than Planck (as do most other probes of this parameter combination).

One method for measuring \fsig\ involves regressing the observed peculiar velocities (from e.g. supernovae, Tully-Fisher or Fundamental Plane standard candles/rulers) on their predicted peculiar velocities from the density field of galaxies, obtained from a redshift survey. Specifically, the slope of this regression, $\beta_g$, is then combined with a measurement of the fluctuations of galaxies, $\sigma_{8,g}$ to yield an estimate of
\begin{equation}
    f\sigma_8 = \beta_g \sigma_{8,g}\,,
\label{eq:estimator}
\end{equation}
as first suggested by \cite{pike2005cosmological}. The background theory and assumptions implicit in this method are discussed in more detail in Section \ref{sec:LPT} below.

The goal of this paper is to test the accuracy and precision of this method using large cosmological $N$-Body simulations at a redshift of zero of DM halos as a proxy for galaxies, as well as semi-analytical models of galaxy formation. Our approach is not to simulate all the observational properties of the surveys simultaneously (as in, for example, \cite{NusDavBra14}), but rather to consider one by one the different physical effects which may bias or add uncertainty to our results. In this sense the paper is similar in spirit to \cite{berlind2000biased}.

This paper is organized as follows. 
Section \ref{sec:LPT} describes the theoretical framework of the relationship between peculiar velocities and the density fluctuation field from linear perturbation theory, and the relevant cosmological parameters.
Section \ref{sec:sim_data} presents the simulation data and semi-analytic models that are used for the analyses performed in this paper.
Section \ref{sec:methodology} describes the prescription for how we predict peculiar velocities. 
In section \ref{sec:SL} we investigate how the choice of smoothing kernel impacts our estimates of \betaf\ and the amount of scatter generated in velocity -- density cross-correlations.
Section \ref{sec:dexes} investigates how uncertainties corresponding to 0.1 and 0.2 dex in halo mass measurements influence the predictions of \betaf\ and $\sigma\sbr{8,h}$. 
In section \ref{sec:galaxies} we instead explore how using either the stellar to halo mass relation or galaxy observables to weight the density field impacts these cosmological estimates.    Section \ref{sec:finite} focuses on how these estimates are affected by volume limited surveys.
Finally, Section \ref{sec:discussion} presents our conclusions.

\section{Peculiar velocities from linear perturbation theory}\label{sec:LPT}

In linear perturbation theory, it is possible to relate the density field to the peculiar velocities of the galaxies at low redshift using
\begin{equation}\label{eqn:v_r}
    v(\boldsymbol{r}) = \frac{H_0 f(\Omega_m)}{4 \pi}\int \delta(\boldsymbol{r}') \frac{ (\boldsymbol{r}'-\boldsymbol{r})}{|\boldsymbol{r}'-\boldsymbol{r}|^3} d^3\boldsymbol{r}'\,,
\end{equation}
where $v(\boldsymbol{r})$ is the peculiar velocity field and $\delta(\boldsymbol{r})$ is the matter density fluctuation field given by
\begin{equation}
    \delta(\boldsymbol{r})=\frac{\rho(\boldsymbol{r})-\bar{\rho}}{\bar{\rho}}\,,
\end{equation}
where $\rho(\boldsymbol{r})$ is the matter density field and $\bar{\rho}$ is its cosmic average.
This calculation is only valid in the linear regime, where $\delta \lesssim 1$, allowing higher order terms to be ignored \citep{peebles1993principles}. For example, it does not predict the transverse components of a galaxy within a galaxy group. In $\Lambda$CDM, the rms matter density in spheres increases with decreasing sphere radius, so we expect linear theory to break down on small scales. In practice then, to apply equation \ref{eqn:v_k} one needs to smooth $\delta(r)$.

In the linear regime, the density modes in Fourier space grow independently of one another. As a result it is easier to write equation \ref{eqn:v_r} in Fourier space as follows
\begin{equation}\label{eqn:v_k}
    \bm{v_k}=iH_0f \frac{\bm{k}}{|\bm{k}|^2}\delta_k\,,
\end{equation}
where $H$ is the Hubble constant ($H = 100\,h\,\mathrm{km}\,\mathrm{s}^{-1} \mathrm{Mpc}^{-1}$). The smoothing is also simpler in Fourier space, because a convolution is a multiplication in Fourier space.
   
The density fluctuation field used in the previous equations is that of the underlying matter density field. Because it is dominated by dark matter, this cannot not be measured empirically. Instead, an assumption must be made as to how the observed galaxies trace the underlying total matter. If one assumes linear biasing, the relation is
\begin{equation}
    \delta_g = b_g \delta,
\end{equation}
where $b_g$ is the linear bias and $\delta_g$ is the density fluctuation field of the \emph{galaxies}. Under this assumption equation \ref{eqn:v_r} can be written as
\begin{equation}\label{eqn:v_r_b}
    v(\boldsymbol{r}) = \frac{H_0 f(\Omega_m)}{4 \pi b_g}\int \delta_g(\boldsymbol{r}') \frac{(\boldsymbol{r}'-\boldsymbol{r})}{|\boldsymbol{r}'-\boldsymbol{r}|^3} d^3\boldsymbol{r}' . 
\end{equation}
Note that if we express distances $\bm{r}$ in units of \h\ Mpc (or km/s), as are naturally obtained from redshift surveys, then one must set $H = 100\,\mathrm{km}\,\mathrm{s}^{-1} \mathrm{Mpc}^{-1}$ (or 1, respectively) in the above expression. Thus when applying this equation to density fields derived from redshift surveys, the predictions are independent of the true value of $H$.
The other two values outside the integral can be compacted into the parameter combination
\begin{equation}
\beta_g \equiv \frac{f}{b_g} \,.
\label{eqn:beta}
\end{equation}
If linear biasing holds, then $\sigma_{8,g} = b_g \sigma_8$. Putting this into equation \ref{eqn:beta}, we find that the product of the observables to be $\beta_g \sigma_{8,g}$, as in equation \ref{eq:estimator}, and so we can set constraints on cosmological parameters.

In reality, the assumptions of linearity, both in the context of perturbation theory and in the context of biasing, will not hold exactly. Fortunately, peculiar velocities are primarily generated by large scale waves, where linearity will be a good approximation. Nevertheless, they also have a contribution from smaller, less linear scales. For these reasons, simulations are needed to calibrate any biases arising from non-linearities.

\section{Simulation Data}\label{sec:sim_data}

We use two publicly available simulations: Bolshoi and MultiDark Planck 2 (MDPL2) \citep{Klypin2011,Klypin2016MultiDark}, along with two simulated semi-analytic catalogues, SAG \citep{Cora2018} and SAGE \citep{Croton2016} which populate the DM haloes of MDPL2. We use the snapshots at $z=0$. The halo and galaxy catalogues were obtained from the COSMOSIM database \footnote{\url{www.cosmosim.org}}.

The  high-resolution Bolshoi simulation \citep{Klypin2011} follows  2048$^3$ particles in a comoving, periodic cube of length 250 \h\ Mpc from $z=80$ to today. It has a mass and force resolution of $1.35\times10^{8} h^{-1} M_{\odot}$ and 1 \h\ kpc respectively, and the DM haloes range from the masses of MW satellites ($10^{10} M_{\sun}$) to the largest of clusters ($10^{15} M_{\sun}$).  It was run as a collisionless  DM  simulation  with  the  Adaptive  Refinement  Tree  Code (ART; \cite{Kravtsov1997}) and assumes a flat, WMAP5 cosmology with parameters $\Omega_m =0.27$, $\Omega_\Lambda=0.73$, $h=0.7$, (linear) $\sigma_8=0.82$ and $n_s=0.95$.  Haloes in Bolshoi were identified using Rockstar \citep{Behroozi2013Rockstar}. The (non-linear) $\sigma_{8,m}$, measured from the particles is 0.897.

The MultiDark  project consists of a suite of cosmological hydrodynamical simulations \citep{Klypin2016MultiDark}, all assume a flat $\Lambda$CDM cosmology with cosmological  parameters:  $\Omega_M=0.307115$, $\Omega_\Lambda=0.692885$, $h=0.6777$, linear $\sigma_8=0.8228$, and $n_s=0.96$, which is consistent with Planck results \citep{planck_cmb_cosmo}. We focus on the MDPL2 simulation which has a periodic box of length 1000 h$^{-1}$ Mpc evolved from a redshift of 120 to 0 with a varying physical force resolution level from 13-5 \h\ kpc and various implemented physics. The simulation uses 3840$^3$ dark matter particles of mass $1.51\times 10^9$ \h \Msol, and has identified more that 10$^8$ haloes using Rockstar \citep{Behroozi2013Rockstar}, with merger trees that were generated using ConsistentTrees \citep{Behroozi2013Trees}. The (non-linear) $\sigma_{8,m}$, measured from the particles is 0.95.

The  SAG \citep{Cora2018} and SAGE \citep{Croton2016} semi-analytic models include the most relevant physical processes in galaxy formation and evolution, such as radiative cooling,  star  formation, chemical  enrichment, supernova feedback and winds, disc instabilities, starbursts,  and  galaxy  mergers. These model were calibrated to generate galaxy catalogues using the MDPL2 simulation. A comprehensive review of the models can be found in \cite{Knebe2017}.

\section{Testing Methods with N-Body Simulations}\label{sec:methodology}

Our goal is to test equation (\ref{eqn:v_r_b}) using data from N-body simulations. Specifically, we will calculate the predicted peculiar velocities by integrating over a smoothed density tracer field, denoted by the subscript ``t'', used in place of $\delta_g(\bm{r})$ in the integral,
\begin{equation}
\bm{v}\sbr{pred}(\bm{r}) = \frac{H_0}{4 \pi}\int \delta_t(\boldsymbol{r}') \frac{(\boldsymbol{r}'-\boldsymbol{r})}{|\boldsymbol{r}'-\boldsymbol{r}|^3} d^3\boldsymbol{r}' \,,
\label{eq:v_r_t}
\end{equation}
and compare this to the ``observed'' (unsmoothed) velocity tracers, which are used in place of $\bm{v}(\bm{r})$ on the left hand side of  equation (\ref{eqn:v_r_b}). Note that equation (\ref{eq:v_r_t}) omits the $\beta_t$ term, which we fit to the N-body data by performing a linear regression
\begin{equation}
\bm{v}_t = \beta_t \bm{v}\sbr{pred}\,,
\end{equation}
where $\bm{v}_t$ is the measured N-body velocity of a tracer of the velocity field which, in principle, need not be the tracer as used for the density field).
This is a cross-correlation between two samples, and we will refer to it generically as a velocity -- density cross-correlation, although the actual comparison is made between observed and predicted peculiar velocities.

The procedure used to obtain the density tracer field is the same regardless of whether the tracer data are particles, DM haloes or galaxies, as each simulation provides Cartesian positions and velocities for all tracers.  The latter two also provide additional information such as halo mass, stellar mass and luminosity measurements in different filter bands.  In this paper, we will construct the density fluctuation field using all of the aforementioned tracers. In the following section however, we will focus solely on using the particles for the density field.

Likewise, one also has a choice of which velocity tracers to use in the comparison. Using the particles as the velocity tracers is the most straightforward, so we will begin with this case in Section \ref{sec:particle-weighting} below. Generally, however observers usually combine the peculiar velocities from multiple galaxies in the same group or cluster to obtain the mean peculiar velocity of the group. The $N$-body analogue for the group-averaged peculiar velocity is the peculiar velocity of the host (or central) halo.

In order to calculate the density field, the density tracers are placed on a 3D cubic grid at the nearest grid point and the contribution of all tracers at the same grid point is summed.

In the case where grid spacing is non-negligible, the gridding acts like smoothing, and it adds in quadrature with the applied to yield a total effective smoothing \citep{BorHudLav20},
\begin{equation}
    \sigma\sbr{total}^2=\sigma\sbr{grid}^2+\sigma\sbr{smooth}^2,
\end{equation}
where $\sigma\sbr{grid}^2 = \frac{\Delta\sbr{grid}^2}{12}$ where $\Delta\sbr{grid}$ is the grid spacing in \h Mpc. 

Using a fine grid allows us to preserve some of the detail and ignore the effects of grid size smoothing. 
All comparisons in this paper are done assuming a grid spacing of 0.36  and 0.98 h$^{-1}$Mpc for Bolshoi and MDPL2 respectively, which produces a negligible effect on the effective smoothing scales that will be used in this paper. 

We choose to smooth the density fluctuation field using a Gaussian smoothing kernel, which,
in configuration and Fourier space, are given by
\begin{eqnarray}
    W(r) & = & \frac{1}{\sqrt{2 \pi} R_G} \exp\left(\frac{-r^2}{2 R_G^2}\right) \\
    W(k) & = &  \exp\left(\frac{-k^2 R_G^2}{2}\right)\,,
\end{eqnarray}
respectively.
We then calculate the peculiar velocities on the 3D grid using equation (\ref{eqn:v_k}). To predict the peculiar velocity of a given velocity tracer, we linearly interpolate from the velocity grid to the tracer's location. 

As discussed above, we then regress the tracer velocity against its predicted velocity, with the intercept fixed to zero, and where the fitted slope gives an estimate for $\beta_t$. The fits also yield the rms (1D) velocity scatter, $\sigma_v$, of the difference between the $N$-body velocities and the linear theory predictions at the best fit $\beta_t$.

\section{The Effect of Smoothing Length on the Estimated Cosmological Parameters}\label{sec:SL}

As discussed above, the density tracer field must be smoothed in order to predict the peculiar velocities in linear perturbation theory. In this section, we assess the impact of different Gaussian smoothing lengths on the recovered value of $\beta_t$, and the 1D $\sigma_v$ around the linear theory predictions.  We focus on smoothing scales between 1-6 \h\ Mpc. 

Specifically, we aim to confirm at what smoothing length the slope of linear regression is unbiased, which previous work has suggested lies in the range 4 -- 5 \h\ Mpc \citep{carrick2015cosmological}. We first consider particles as tracers of the density field, and then consider haloes.

\subsection{Particle-weighted Density Fluctuation Field}\label{sec:particle-weighting}

\begin{figure}
    \centering
    \includegraphics[width=.5\textwidth]{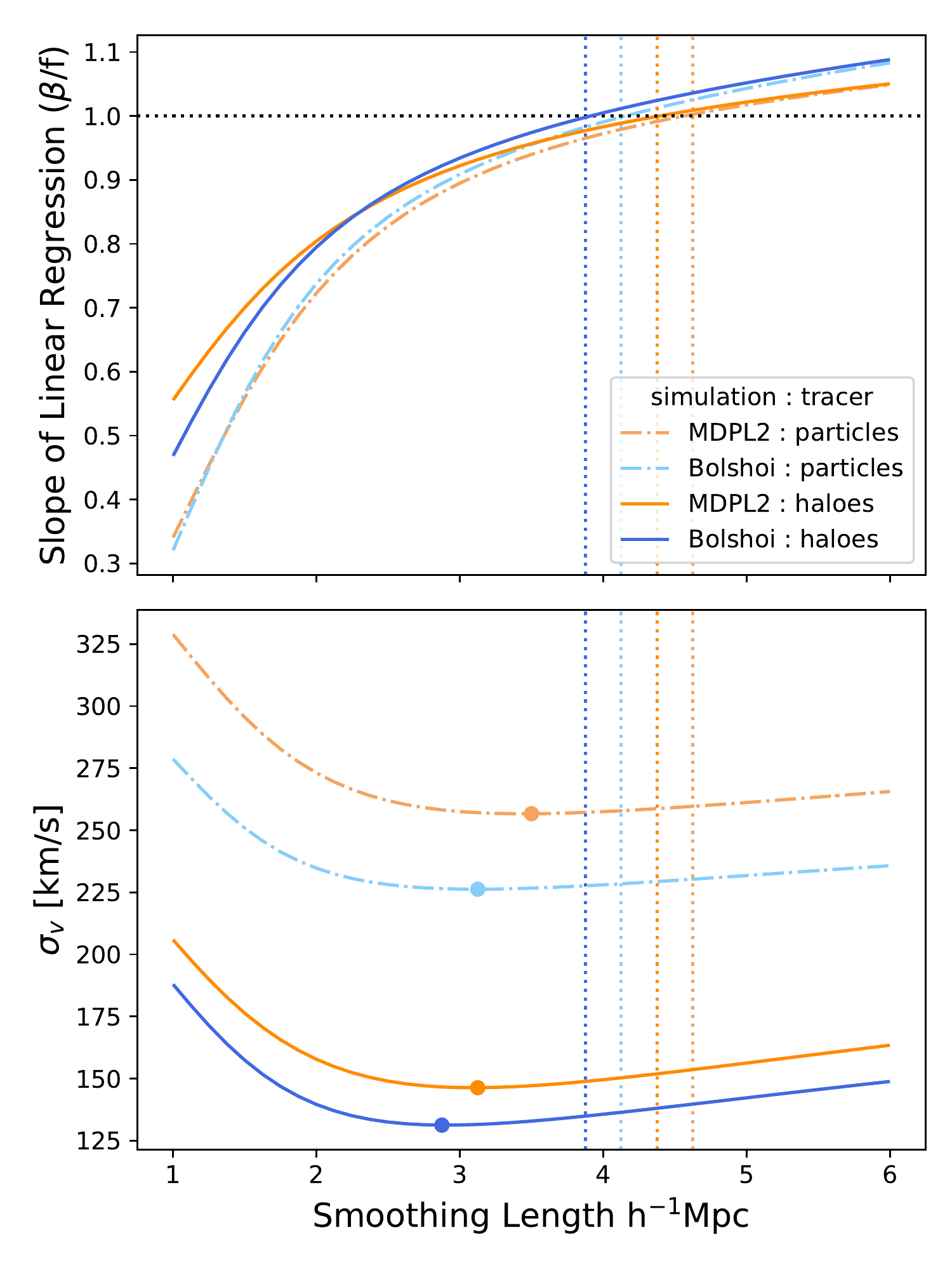}
    \caption{Top Panel: Slope of linear regression for the predicted and N-Body particle and central halo velocities (excluding subhaloes) for MDPL2 and Bolshoi simulations. Because we correct for the $f$ parameter, a value of $\beta/f$ of unity indicates for the smoothing length for which the velocity--density cross-correlation is unbiased. The predicted velocities were calculated assuming an underlying \emph{particle} density field and linear theory for the Gaussian smoothing radius shown on the horizontal axis . Bottom Panel: The scatter between the measured $N$-body and predicted peculiar velocity associated with each of the linear regression slopes. A circle has been placed at the smoothing length where the standard deviation was minimized.}
    \label{fig:part_part_sl}
\end{figure}

We now consider the case in which we predict the peculiar velocities of particles using the particle density fluctuation field. The values of the best-fit slope divided by the value of $f$ in the simulation, i.e. $\beta/f$, as a function of smoothing scale, are shown in  Figure \ref{fig:part_part_sl}. This quantity should be unity if the method is unbiased. We find that predictions are unbiased for a Gaussian smoothing kernel $R_G$ between 4 and 5 \h Mpc. Of the two simulations, MDPL2 should be more accurate since its minimum wavenumber $k$ is $2\pi \times 10^{-3} h/$Mpc, and whereas that of Bolshoi is a factor 4 larger due to its smaller box length. Thus Bolshoi fails to capture the long-wavelength modes that generate a significant fraction of the rms peculiar velocity.  The 1D $\sigma_v$ around the best fit slope, however, is minimized at a smoothing length that is 1-1.5 \h\ Mpc smaller. The $\sigma_v$ of particle peculiar velocities is high: between 225 and 275 km/s. This is because the particle's peculiar velocity includes its motion within the halo as well as the peculiar velocity of the halo itself. Only the latter is well-predicted by linear perturbation theory.
    
If we predict the peculiar velocities of host (or central) haloes (i.e.\ excluding subhaloes) using the particle density field, we find similar results, with unbiased results and minimum $\sigma_v$ occurring for Gaussian smoothing lengths that are $\sim 0.5 \h $Mpc smaller than when using particles as velocity tracers. The  $\sigma_v$ as a function of smoothing length is quite flat; it is not much higher at the fiducial 4 \h\ Mpc. However the amplitude of $\sigma_v$ for host haloes is significantly lower ($\sim 150$ km/s) than for particles ($\sim 250$ km/s), as expected given that the velocities of the particles include their motion with respect to the host halo.  

\begin{figure}
    \centering
    \includegraphics[width=.5\textwidth]{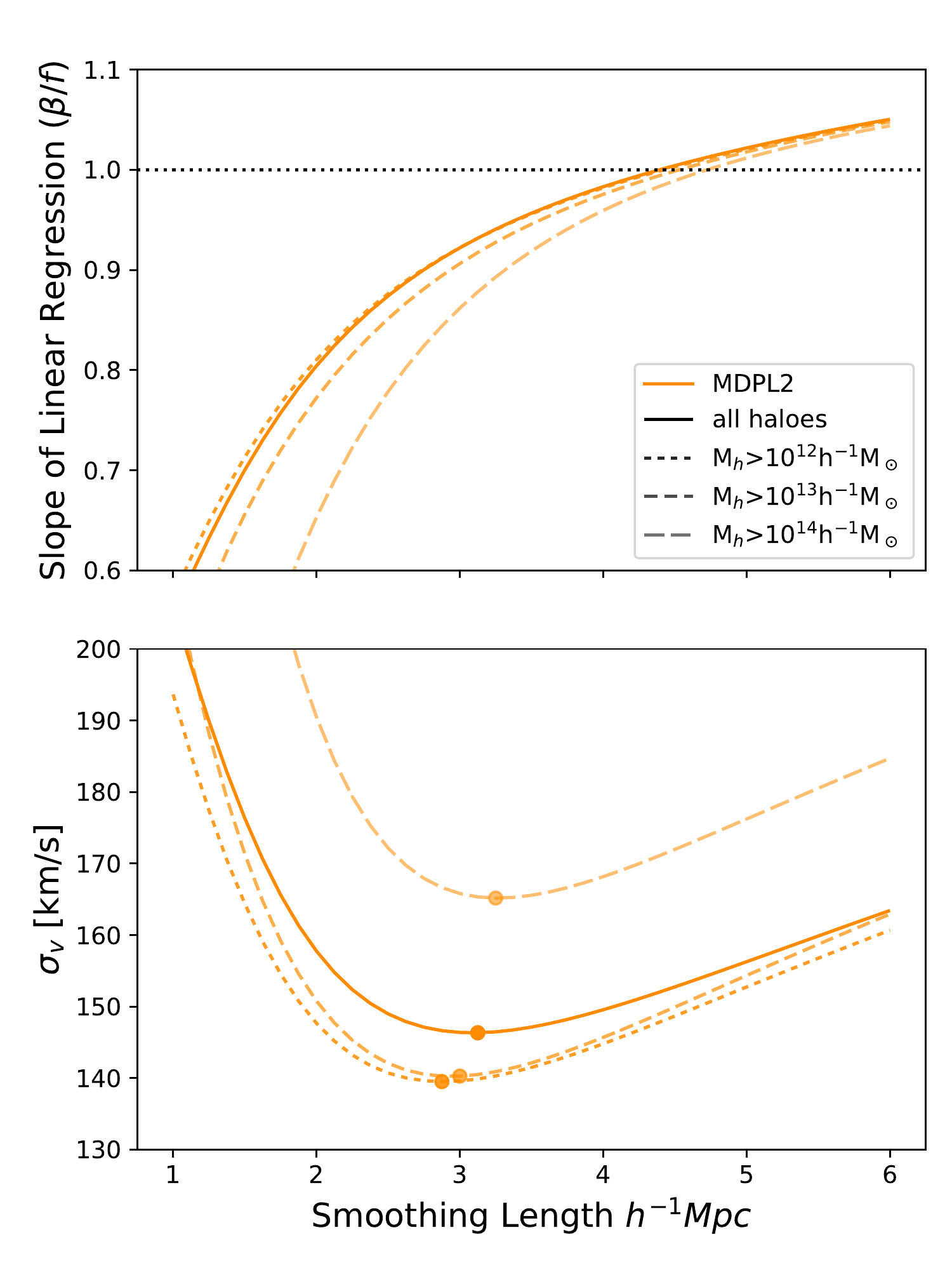}
    \caption{As in Fig. \ref{fig:part_part_sl} where again the predicted peculiar velocities are based on the particle density field, but here these are compared to the $N$-body peculiar velocities of \emph{haloes} (but excluding subhaloes). For clarity the solid line labeled 'all halos' represents all simulation objects classified as centrals, while the dashed lines represent centrals with masses greater than various minimum masses as indicated in the legend.}
    \label{fig:particles_sl}
\end{figure}

Finally, it is interesting to explore the question of whether haloes of different masses have different velocities relative to the predictions of linear theory, one scenario of ``velocity bias.'' 
We find that imposing a minimum mass threshold on the haloes used to sample the velocity field of $10^{12}$ \Msol\ has little effect on the measured $\beta/f$ for smoothing lengths greater that 1.5 \h Mpc for MDPL2 and 3 \h Mpc for Bolshoi, as shown in Fig.\ \ref{fig:particles_sl}. This is also true for a minimum mass of $10^{13} \Msol$. A small velocity bias appears only when considering cluster mass haloes ($>10^{14} \Msol$) as peculiar velocity tracers. For clusters, there is also an increase in the 1D $\sigma_v$ with respect to the linear theory predictions.

\subsection{Halo mass-weighted density field}\label{sec:halo-weighting}

\begin{figure}
    \centering
    \includegraphics[width=.5\textwidth]{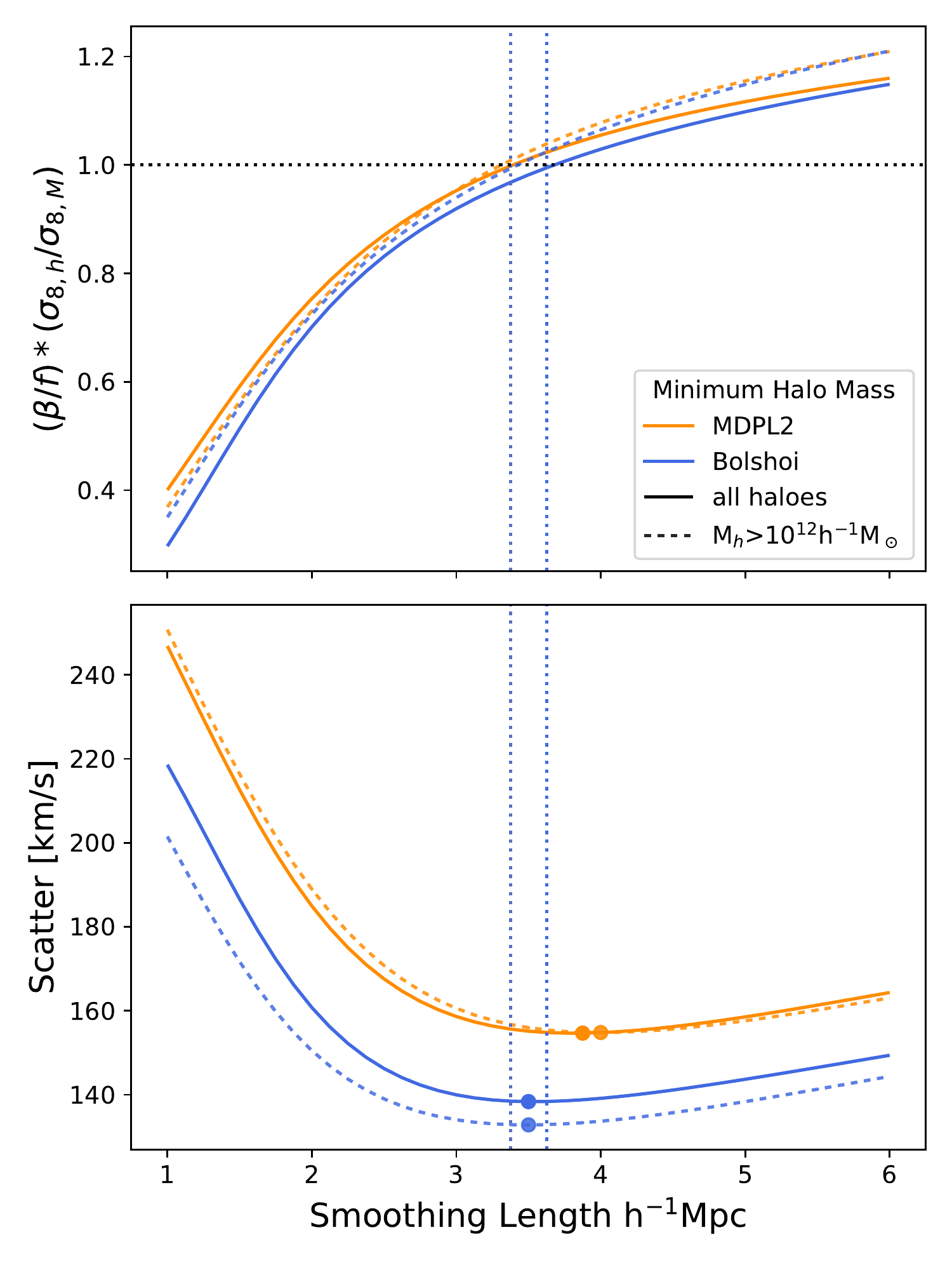}
    \caption{Similar to Fig. \ref{fig:particles_sl}, except here the mass-weighted halo density field is used to predict peculiar velocities of haloes. The vertical axis in the top panel is now $\beta/f (\sigma_{8,h}/\sigma_{8,m})$. }
    \label{fig:mass_sl}
\end{figure}

We now consider a scenario that is closer to the observational one, where the density field is provided by DM haloes, weighted by their mass. Whereas the particle density field is unbiased, this field will be biased. Therefore, we no longer expect $\beta\sbr{h}/f = 1$. As discussed above, in the linear regime, this bias can be calculated by measuring the rms density fluctuations of the halo-weighted density field in 8 \h\ Mpc spheres. The halo and particle $\sigma_8$ measurements are related by 
\begin{equation}
    b\sbr{h} = \frac{\sigma\sbr{8,h}}{\sigma\sbr{8,m}}
\end{equation}
which is the same $b$ in the $\beta$ term defined previously. We measure $\beta/f$ from the slope of linear regression as before, but multiplying this by the correction factor $b\sbr{h}$. This should return a value of unity if the method unbiased. This parameter combination will be referred to as $\beta^* =  \frac{\beta}{f} \frac{\sigma\sbr{8,h}}{\sigma\sbr{8,m}}$.

To measure \sig\, we placing non-overlapping spheres of 8 \h Mpc covering the entire the density field, and measure the standard deviation in the values.
Doing this, we find values of $\sig\sbr{h} = 1.45 \pm 0.04$ and $1.57 \pm 0.02$ for the halo masses for Bolshoi and MDPL2 respectively, much larger than $\sigma\sbr{8,m}$, which is the measured (non-linear) \sig of the particle density field.

The fitted values of $\beta\sbr{h}/f$, however, are lower than unity, as expected. The results for $\beta^*$ are plotted in Figure \ref{fig:mass_sl}, showing that the mass-weighted determination is nearly unbiased: at the fiducial $R_G = 4 \h$ Mpc, the MDPL2 field has $\beta^* = 1.05$, whereas for Bolshoi $\beta^* = 1.02$. This suggests that linear biasing correction works well, even for fields with $b\sbr{h} \sim\ 1.7$. The 1D $\sigma_v$ ($\sim 160$ km/s) is only marginally higher than when particles were used for the density field.

\subsection{Discussion: Cross-Correlation and Optimal Smoothing}\label{sec:cross-corr}

It is not obvious why a Gaussian smoothing, with $R_G \sim 3-4$ \h\ Mpc, should be the ``correct'' smoothing length. We can gain some insight by considering the problem in Fourier space, specifically equation \ \ref{eqn:v_k} which relates the velocity modes to the density modes in linear perturbation theory.  This should be exact on large scales, but we expect it to break down at high $k$. 

For simplicity, first assume that all Fourier modes are Gaussian and independent and that at a given $k$ the joint distribution of velocity and density mode amplitudes is given by a bivariate Gaussian. We expect the correlation coefficient, $\rho$, to approach 1 on large scales and zero at high $k$. Generally if one has a bivariate normal distribution of two variables $x$ and $y$, and one regresses $y$ on $x$ the slope of the relation is $\rho \frac{\sigma_y}{\sigma_x}$, or since the covariance $C_{xy} = \rho \sigma_x \sigma_y$, the slope can also be written $C_{xy}/\sigma_x^2$. Therefore, the predicted velocity mode amplitude is given by the density mode amplitude times the slope $C_{v\delta}/\sigma_\delta^2$. 

\begin{figure}
    \centering
    \includegraphics[width=.5\textwidth]{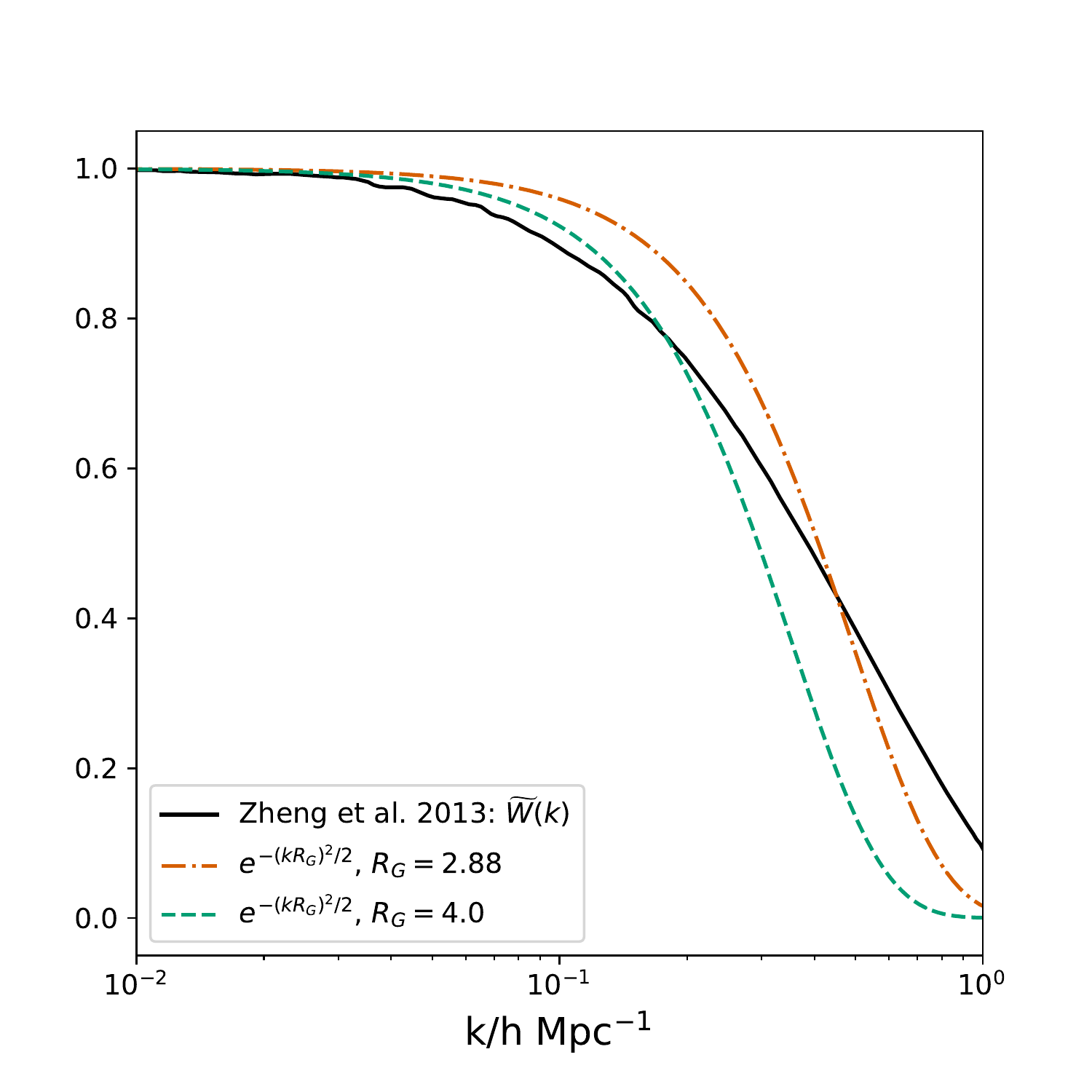}
    \caption{The normalized window function from \protect\cite{Zheng2013}, calculated from the J1200 simulation, based on the correlation of the velocity and density fields (see text for details) is shown, with Window functions for Gaussian smoothing kernels with scales of 2.85 and 4.0 \h\ Mpc. Both functions exhibit the expected characteristics of being unity as $k \rightarrow 0$ and zero as $k \rightarrow \infty $.  $\widetilde{W}_k$ and the Gaussian kernel of $R_G=2.88$ \h\ Mpc cross 0.5 near $k = 0.39$ and $0.41 h/$Mpc, respectively.} 
    \label{fig:Wk}
\end{figure}

The quantities $C_{v\delta}$ and $\sigma_\delta^2$ can be measured in $N$-body simulations as a function of wavenumber $k$. The latter is just the power spectrum of matter density fluctuations, $P_{\delta \delta}$, but the covariance is less well studied. \cite{Zheng2013} have measured the cross-spectrum of the closely related quantity $\theta = \nabla \cdot v$ with the density $\delta$, and its ratio with the density power spectrum. They define the normalized window function 
\begin{equation}
    \widetilde{W}_k = \frac{1}{f} \frac{P_{\theta \delta}}{P_{\delta \delta}} \,,
\end{equation}
which has the property that it asymptotes to $1$ at low $k$ and goes to zero at higher $k$. This function is plotted in Figure \ref{fig:Wk}. Also overplotted for comparison are two Gaussian smoothing filters with $R_G = 2.88$ and 4.0\h\ Mpc.  The estimated \betaf\ from the particle weighted density field smoothed using either $\widetilde{W}_k$ as a $k$-space smoothing function or with a Gaussian kernel of $R_G = 2.88$ \h Mpc both produce values of 0.88. There is also no significant difference in $\sigma_v$.

The Gaussian smoothing with $R_G = 4.0 \h$ Mpc is a better match to $\widetilde{W}_k$ at low $k$.

In reality, particularly at high $k$, the assumption of bivariate Gaussianity will no longer be correct, so the above simple argument will break down. In principle, with detailed knowledge of the correlations, it should be possible to design the optimal $k$-space filter. For our purposes in this paper, we retain the simplicity of Gaussian smoothing, and adopt a Gaussian smoothing filter with $R_G = 4$ \h Mpc for consistency with previous work.

\section{The effect of uncertainty in the halo masses}\label{sec:dexes}

\begin{figure}
    \centering
    \includegraphics[width=.5\textwidth]{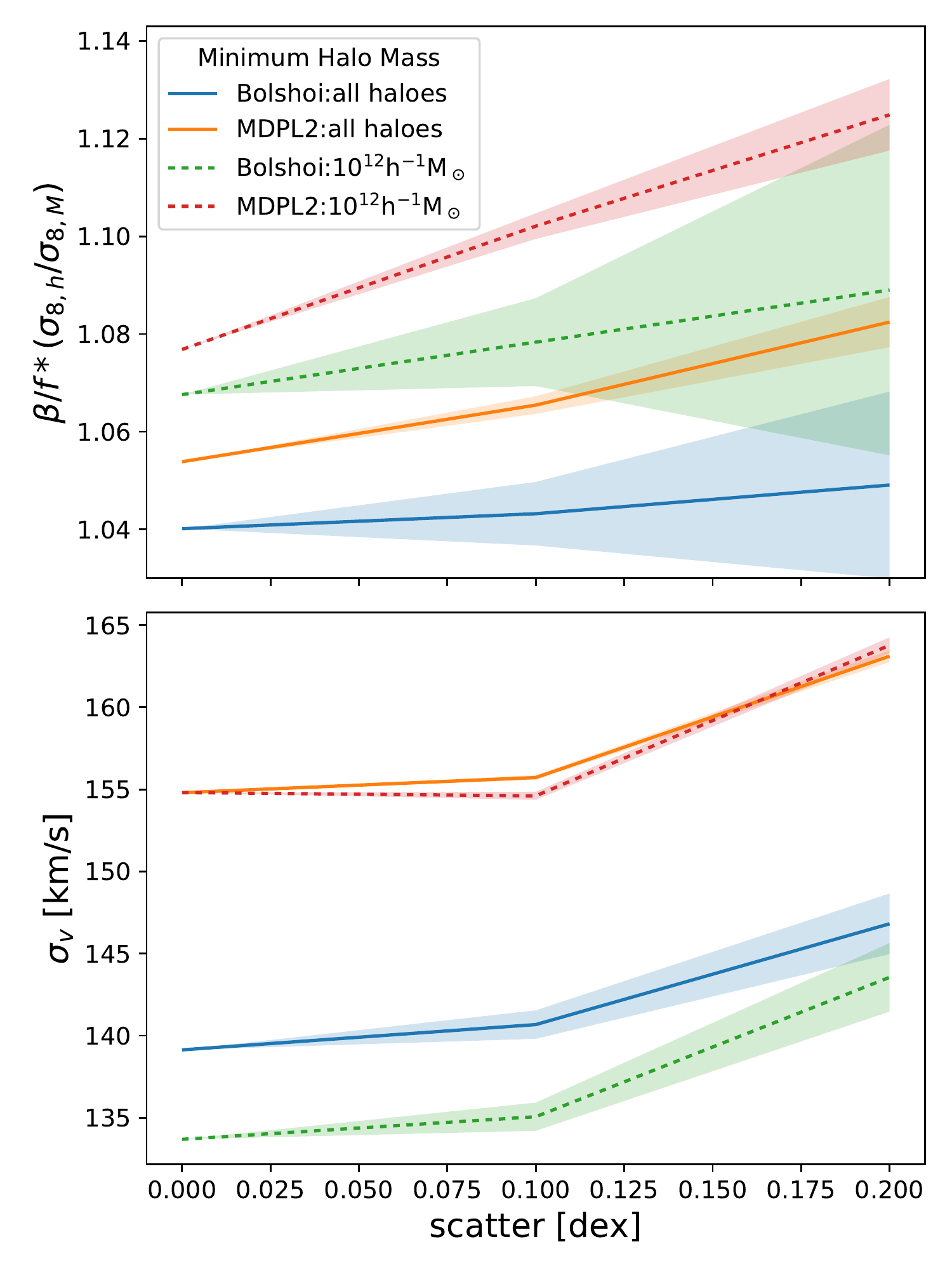}
    \caption{Effect of additional scatter to the mass of haloes as a function of the logarithmic scatter (0, 0.1 and 0.2 dex). For 0 dex this is the measured value of the simulation, for 0.1 and 0.2 dex a random value corresponding to a lognormal Gaussian with a $\sigma$ of $\sim$ 0.23 and 0.5 respectively, has been applied to each halo in the catalogue. Top panel: The mean (dark line) and $\pm 1\sigma$ range (shaded band) of these measurements are shown for $\beta/f (\sigma_{8,h}/\sigma_{8,m})$ at a smoothing radius of 4 \h Mpc. Bottom panel : scatter around the best fit of $\sigma_v$ for the same smoothing.}
    \label{fig:slope_dex_comparisons}
\end{figure}

In the previous sections, the masses of the haloes were assumed to be known exactly, with no uncertainty in the measurements. However in real survey data there is a some uncertainty in the true mass of any given halo, depending on the method used to estimate it, and this may be in the range 0.1 -- 0.2 dex (see Section \ref{sec:galaxies}). In this section, we explore how scattering the halo mass impacts the predictions of \betaf\ and  $\sigma_{8,h}$. This is accomplished by varying the halo masses by a lognormal Gaussian random variable with standard deviations corresponding to 0.1 and 0.2 dex, and calculating \sig\ and $\beta$ for each realization. A total of 500 realizations were performed on both the Bolshoi and MDPL2 halo catalogues.

We find that, from realization to realization, both the measured $\beta$ and \sig\ deviate from the no-scatter values, and these deviations are anti-correlated. This can be understood as follows. In Section \ref{sec:cross-corr}, we discussed how the slope ($\beta$) could be expressed as the ratio of the covariance between density and velocity, and the density power spectrum, $C_{v\delta}/\sigma_{\delta \delta}^2$. If noise is added to the density field, then the denominator increases, but the covariance is unaffected. Consequently, the fitted $\beta$ drops. On the other hand, \sig\ increases because of the additional noise. This anti-correlation leads to some cancellation of the additional noise from realization to realization in the product $\bstar = \beta\sbr{h} \sigma\sbr{8,h}$.

The net result of increasing the halo mass scatter and its impact on \bstar\ are shown in the top panel of Figure \ref{fig:slope_dex_comparisons}, with a scatter of zero dex corresponding to the case where the halo masses are precisely measured. We find that when introducing a lognormal scatter,

the value of \bstar\ increases only slightly with increasing scatter for both Bolshoi and MDPL2.
We note however that the standard deviation of \bstar\ is higher for Bolshoi due to the smaller volume of the simulation, which leads to greater variation from realization to realization in $\beta\sbr{h}$, $\sigma\sbr{h}$ and their product \bstar. For MDPL2, the net effect of scatter in the $N$ independent halo masses is reduced by a factor $1/\sqrt{N}$ due to the larger volume containing a larger number $N$ of haloes. 
This highlights the importance of the volume of the sample, a topic we shall discuss in greater detail in Section \ref{sec:finite}. 

The bottom panel shows how the 1D $\sigma_v$ is impacted by introducing stochasticity in the halo masses. We find that the 0.1 dex case generates a negligible change in the measured $\sigma_v$ when compared to the original value. For halo mass uncertainties of 0.2 dex, $\sigma_v$ increases, but the effect is still small, corresponding to $\sim$ 8 km/s. 

\section{Galaxies as tracers of the density field}\label{sec:galaxies}
\begin{figure}
    \centering
    \includegraphics[width=.5\textwidth]{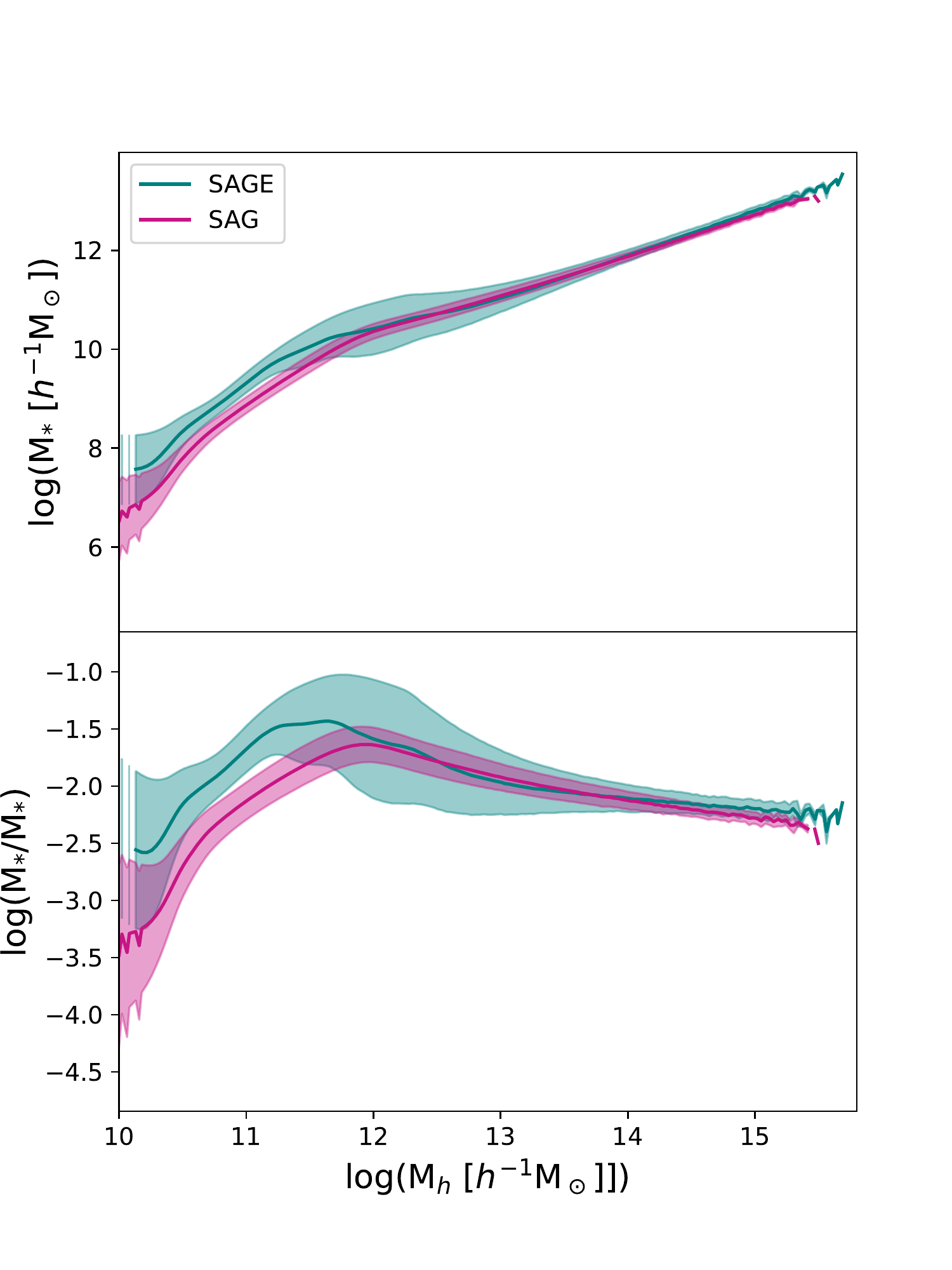}
    \caption{Top Panel: The stellar-to-halo mass relation (SHMR) for the total stellar mass of a halo (including stellar mass in subhaloes) as a function of total halo mass. The dark curves show the means of the SHMR and the lighter bands indicate their measured standard deviation. Both SAG and SAGE semi-analytic models are shown. Bottom Panel: the stellar to halo mass ratio as a function of halo mass.}
    \label{fig:SHMR}
\end{figure}

Galaxies trace the underlying dark matter distribution on large scales, but observable quantities, such as a galaxy's luminosity and stellar mass, do not necessarily have an exact relationship with the mass of the halo in which a galaxy lies.  To explore how using these quantities as a proxy for mass density impact \betaf\ and \sig, we use two galaxy semi-analytic galaxy formation models available for the MDPL2 simulation SAG \citep{Cora2006} and SAGE \citep{Croton2016}. 

Many DM haloes host more than one galaxy, which are typically divided into centrals (the galaxy identified with the main or host DM halo) and satellites (associated with DM subhaloes). When predicting host halo mass, one can use only the stellar mass or luminosity of the central galaxy, or one can use the total stellar mass (or total luminosity) of all galaxies. We adopt the latter approach here
when calculating the density fields. As before, velocity comparisons will be done solely on galaxies classified as centrals.
For consistency with the previous work done in this paper, any cuts imposed on the data will be done using the mass of the host halo.

The stellar to halo mass relation (SHMR) is different for both the SAG and SAGE semi-analytic models in MDPL2, see \ref{fig:SHMR}. SAG has a tighter SHMR with less scatter in stellar mass at a given halo mass:
for haloes with masses between $10^{11}$ and $10^{13}$ \h \Msol, it has an average scatter of 0.15 dex compared to 0.39 dex in SAGE. 
The average SHMR is comparable for both SAG and SAGE for halo masses greater than the characteristic pivot point at $\sim 10^{12}$ \h \Msol.

\subsection{Predictions using Stellar-to-Halo Mass Relations}\label{sec:SHMR_predictions}

\begin{figure}
    \centering
    \includegraphics[width=.5\textwidth]{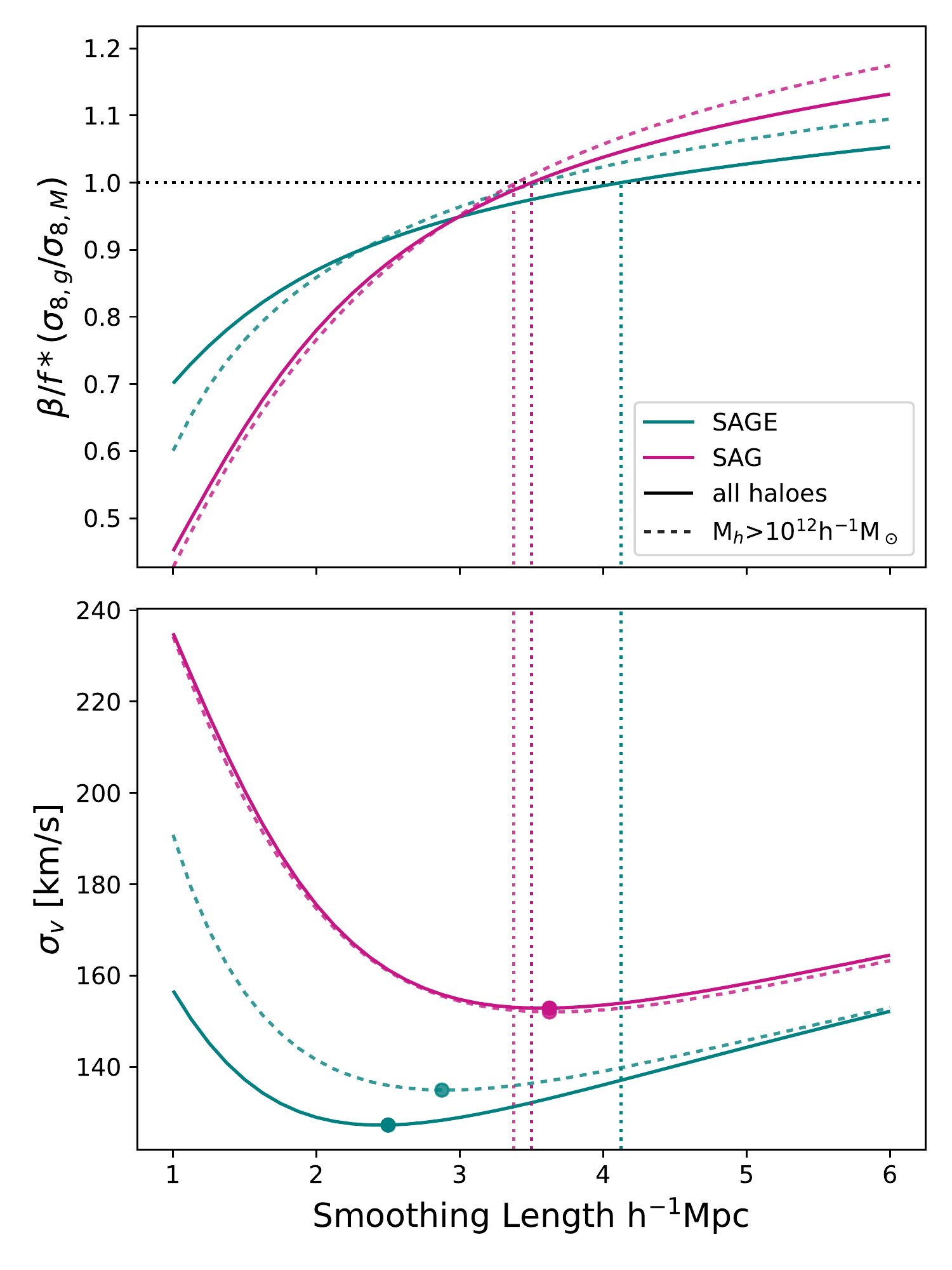}
    \caption{Similar to Fig.\ \ref{fig:mass_sl}, except here a fitted SHMR is used to predict the halo masses from stellar masses, and then constructing a mass-weighted halo density field to predict the peculiar velocities of central galaxies.}
    \label{fig:halo_from_stellar}
\end{figure}

To construct a proxy mass density field, we can obtain proxy halo masses using the SHMR. The K-band luminosity is a good proxy for stellar mass, however the galaxy luminosities in SAG are limited to the $u,g,r,i,z$-bands.  So in this section, we use the stellar masses provided for both SAG and SAGE. 
We fit the SHMR profiles of both SAG and SAGE semi-analytic models with a broken power law, and use these to convert the stellar masses into halo masses. Then the same analysis described in a previous section is performed. This procedure will remove the nonlinearity of the SHMR relation, but cannot remove the scatter in it, since the mean SHMR is used for each halo.

From Figure \ref{fig:halo_from_stellar}, we find for the SAGE galaxies, \bstar\ is unbiased for a Gaussian smoothing of 4.1 \h Mpc but generates the lowest $\sigma_v$ at $R_G =$ 2.6 \h\ Mpc. For SAG these are at smoothing lengths of  3.6  and 3.5 \h Mpc, respectively. Comparing these results to Fig.\  \ref{fig:mass_sl}, we find that at $R_G$ = 4 \h Mpc, SAG produces results that are quite similar (within a few percent) to what was found using MDPL2's halo masses directly. 
SAGE, however, generates values of \bstar\ that are 5\% smaller at the same smoothing length than the haloes. We find that the generated SAG haloes produce similar estimates for the 1D scatter in velocity predictions, contrarily the SAGE haloes generate $\sim$ 20 km/s less scatter. 

Comparing these results to those in the next section we note that while the halo mass does tend to be unbiased at smaller $R_G$ the conversion of stellar to halo mass introduces 10-20 km/s of additional scatter than using galaxy observables to predicted velocities. 

We attribute the results to the differences in the models' stellar to halo mass ratios as a function of halo mass.  SAG has a flatter stellar to halo mass ratio than SAGE, and therefore it is close to the simple halo mass weighted case. SAGE has a steeper ratio at high halo masses, hence SAGE puts less weight on massive clusters leading to estimates of \bstar\ slightly less than unity.

\subsection{Predictions Using Galaxy Observables}\label{sec:observables}

\begin{table*}
\caption{Summary of \bstar values taken for  $R_G = 4$ \h Mpc for the various MDPL2 tracers which weight the density field and from which the peculiar velocities are compared.}
\begin{tabular}{|l|c|c|l|}
\hline
Density Tracer   & \multicolumn{1}{c|}{all M$_{t}$} & \multicolumn{1}{c|}{$M_{t}> 10^{12} \h \Msol $} &\\ \hline
Particles     & 0.97      &     --           & Figure \ref{fig:part_part_sl}      \\ 
Haloes       & 1.05        & 1.08  & Figure \ref{fig:mass_sl}         \\ 
SAGE: SHMR    & 1.00      & 1.02   & Figure \ref{fig:halo_from_stellar}      \\ 
SAG: SHMR     & 1.04       & 1.06   & Figure \ref{fig:halo_from_stellar}      \\ 
SAGE: stellar mass & 0.98      & 1.01  & Figure \ref{fig:pvs_from_observables}        \\ 
SAG: stellar mass  & 1.01       & 1.03  & Figure \ref{fig:pvs_from_observables}        \\ 
SAG: $r$-band luminosity  & 0.97      & 1.02  & Figure \ref{fig:pvs_from_observables}     \\ \hline
\end{tabular}
\label{tbl:beta_4hMpc}
\end{table*}

\begin{figure}
    \centering
    \includegraphics[width=.5\textwidth]{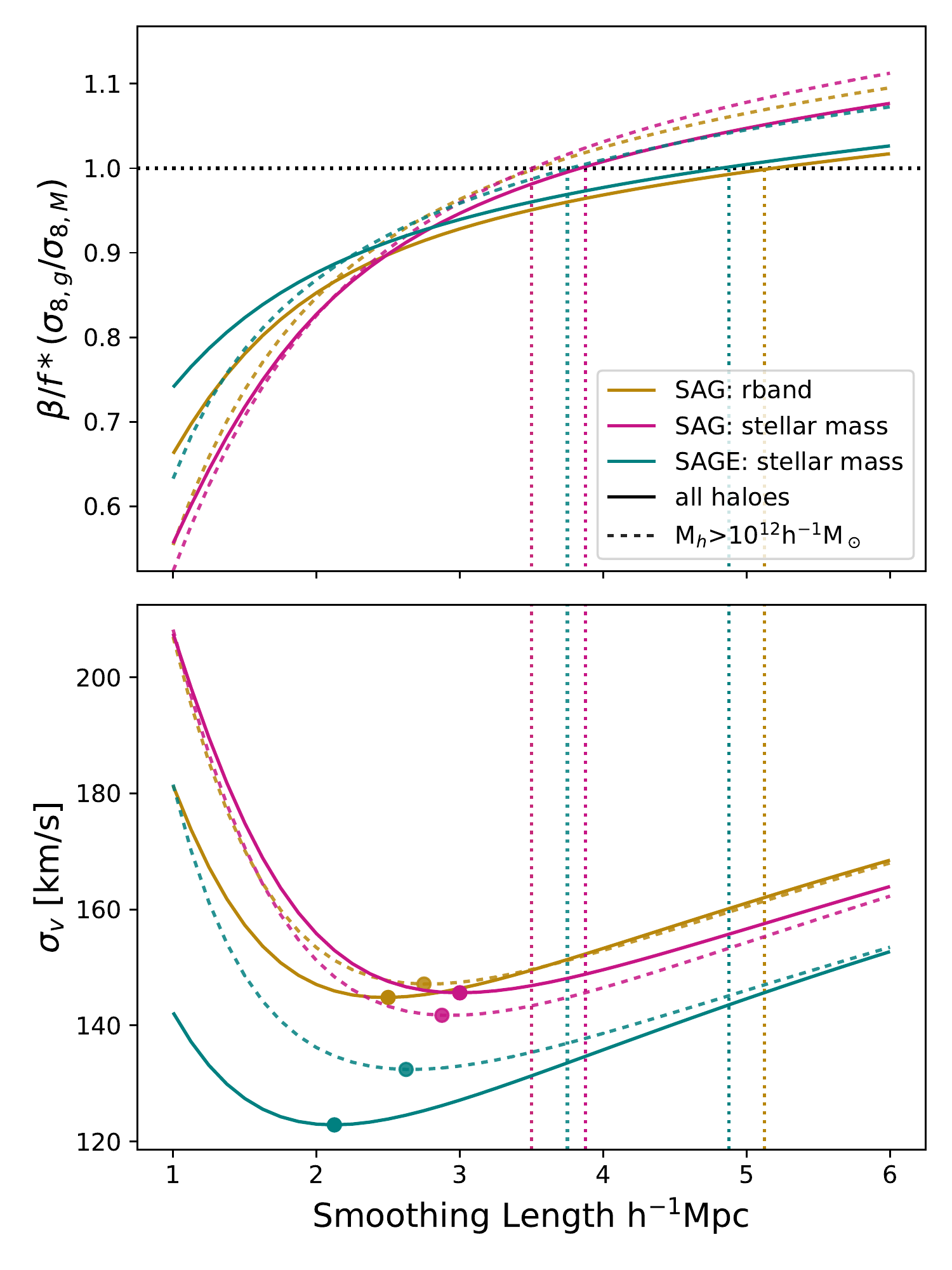}
    \caption{Similar to Fig.\ \ref{fig:halo_from_stellar} except now the density field is constructed using galaxy observables (i.e.\ weighted by stellar mass or luminosity) to predict and measure centrals' peculiar velocities.}
    \label{fig:pvs_from_observables}
\end{figure}

A density field can also be constructed without approximating the mass of individual haloes via the SHMR. In this section, we investigate how weighting directly by the galaxy observables impacts the cosmological estimates. Such a procedure is closer to what was done by \cite{carrick2015cosmological}, who used the $K$-band luminosity-weighted density field from the 2M++ catalogue. Figure \ref{fig:pvs_from_observables} shows the summary of results discussed below.

We find that, for the SAG model, weighting the density field by stellar mass produces results that yield a higher \bstar\ than luminosity weighting. We can weight the density field using luminosities in any of the five bands provided by SAG. In the remainder of this paper we focus on the $r$-band luminosity. We note however that the longest wavelength $z$-band produces the highest values of \betaf\, however after applying the correction factor $(\sigma\sbr{8,g}/\sigma\sbr{8,m})$ there is virtually no difference between bands for  \bstar, provided a low minimum mass threshold.

If the minimum mass threshold of the haloes is low, we find that weighting using the stellar masses provided by SAG closely resembles the case where the density field is halo-mass weighted and is unbiased at an $R_G \sim 3.8$ \h\ Mpc. The SAGE stellar mass and SAG $r$-band predict \bstar\ values that are comparable at $R_G > 3$ \h\ Mpc, but only produce unbiased estimated of \bstar\ for a Gaussian smoothing radius of $\sim$ 5 \h\ Mpc.

In the case where a minimum mass threshold of $10^{12} \h$ \Msol\ is applied to the data, we find that \bstar for the three cases are comparable for $3 \textrm{\h Mpc} \lesssim R_G \lesssim 4 \textrm{\h Mpc}$. With all the cases estimating unbiased \bstar\ values at 3.5-3.7 \h Mpc. The $\sigma_v$ is also comparable for these cases with the SAGE stellar mass producing only $\sim$ 10 km/s less scatter than the SAG $r$-band.

\section{Finite Volume and Cosmic Variance Effects}\label{sec:finite}

\begin{figure}
    \centering
    \includegraphics[width=.5\textwidth]{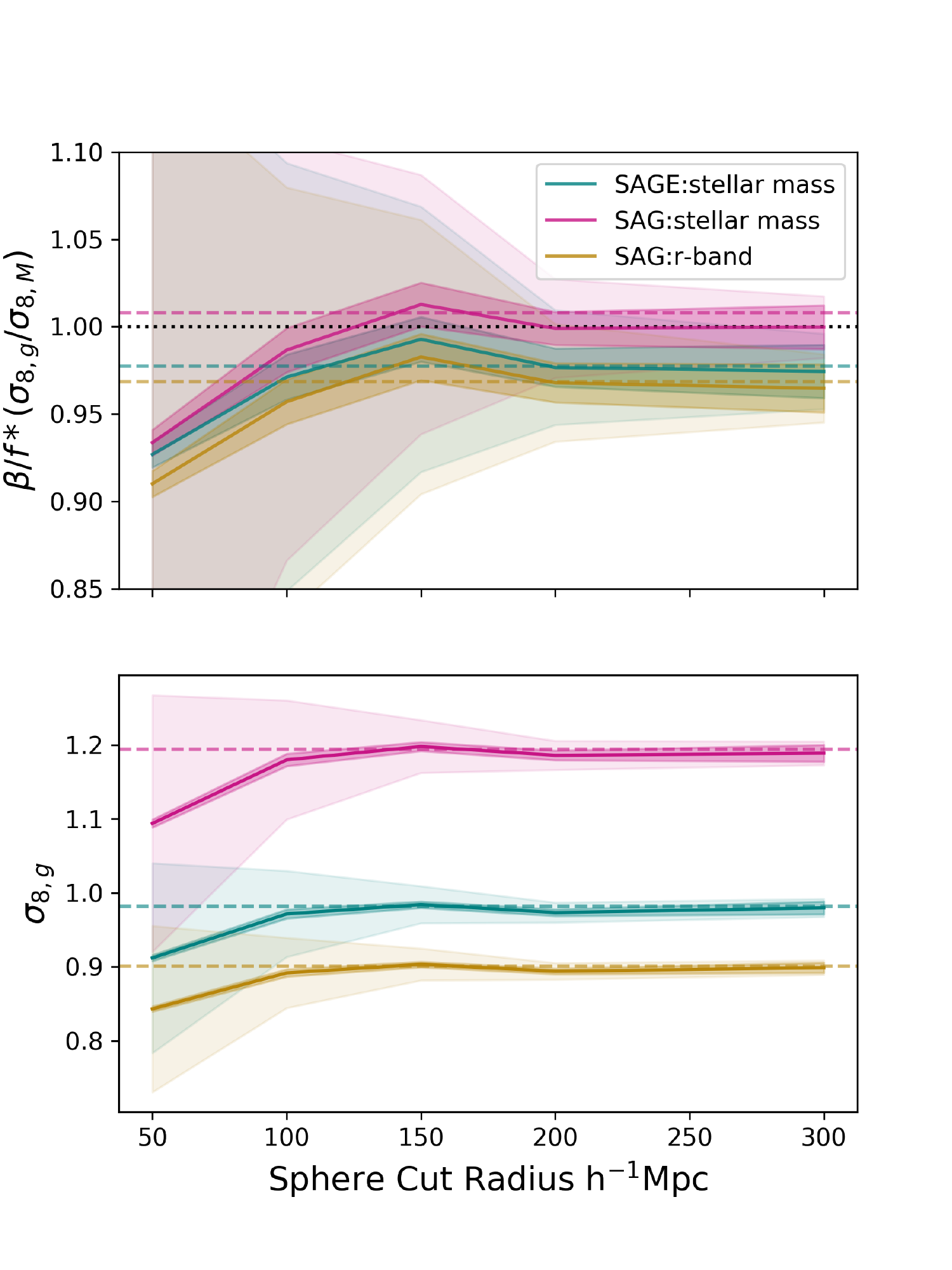}
    \caption{To demonstrate the dependence of our calculated parameters on the survey volume, we restrict simulation data to (non-overlapping) spheres of radius ranging from 50 to 300 \h Mpc. The components of \bstar\ are calculated for each test sample assuming a Gaussian smoothing kernel of 4 \h Mpc.  The upper panel and lower panel show the results for \bstar\ and in \sig\ measurements, respectively, as a function of sphere radius. In both panels, the light and dark coloured bands represent the $\pm 1 \sigma$ standard deviation from sphere to sphere and the standard error in the mean of the test cases (respectively) for a given sample spherical radius, with the mean being shown by the solid colour lines. The colour horizontal dashed lines show the values from the full simulation box.}
    \label{fig:slope_as_fn_sc}
\end{figure}

\begin{figure}
    \centering
    \includegraphics[width=.5\textwidth]{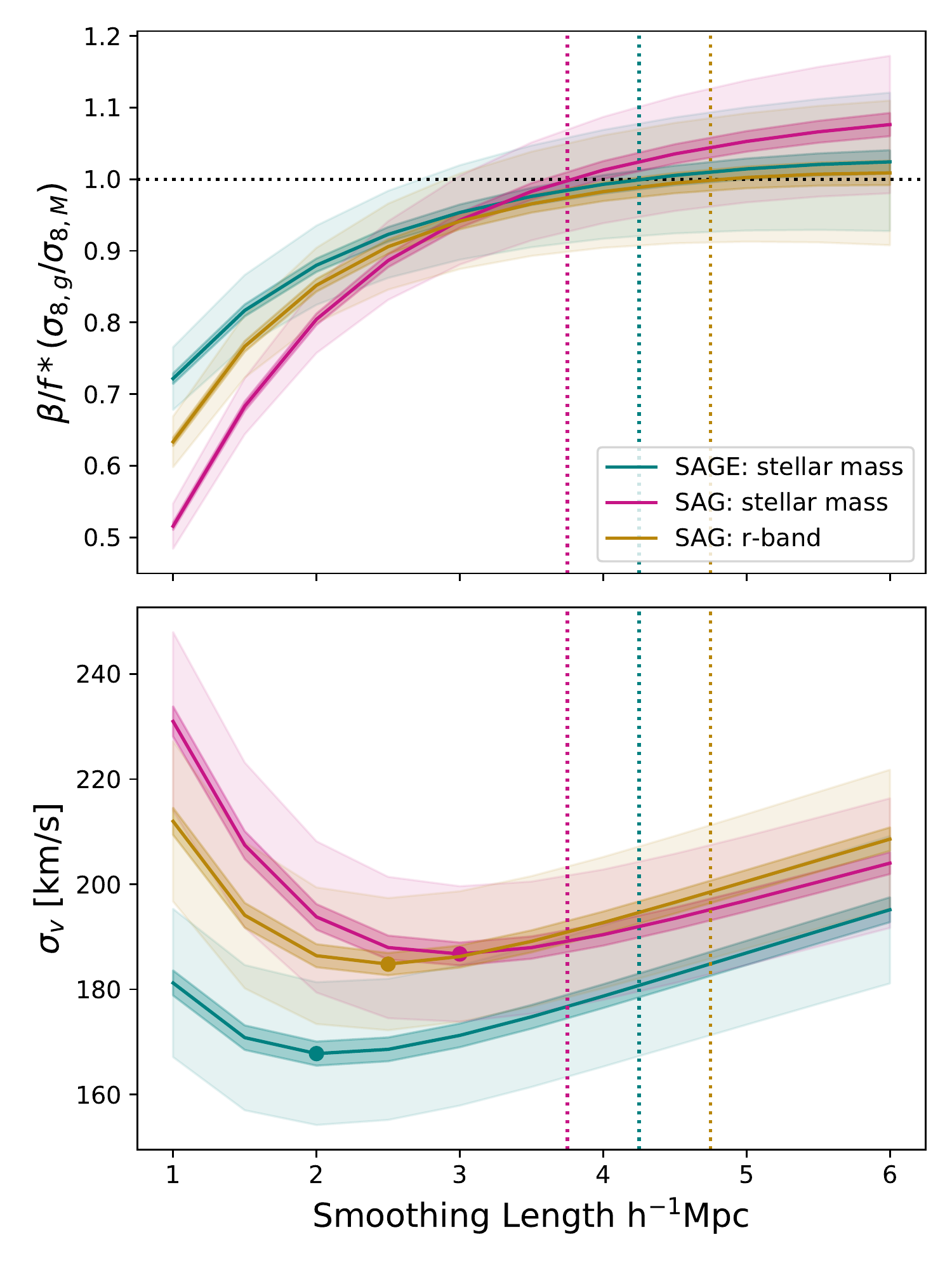}
    \caption{Similar to Fig.\ \ref{fig:pvs_from_observables} expect we now use non-overlapping sphere cuts of radius 150 \h Mpc and the components of \bstar are calculated for a range of Gaussian kernel lengths between 1-6 \h Mpc. The bands have the same meaning as in Fig.\ \ref{fig:slope_as_fn_sc}.}
    \label{fig:sc_150}
\end{figure}

In real data sets both the data used to obtain the density field, based on redshift surveys, and the peculiar velocity samples, cover limited volumes.  This has multiple consequences. Some of these are related to sample variance: the local volume will have local values of the mean density and of \sig\ that are different from the global values. Moreover, the predicted peculiar velocities will be less accurate for galaxies close to the edge of the density field than for those far from the edges, say at the center of the volume.

In this section, we investigate these finite volume and edge effects by limiting the data, to spheres of radii with $R\sbr{max} =$ 50, 100, 150, 200 and 300 \h Mpc, and its impact on cosmological estimates. In particular, we are interested in simulations of data sets  with $R\sbr{max} \sim 150 \h$ Mpc, which is comparable to the survey size of 2M++, the depth of which varies from 125 \h\ Mpc to 200 \h\ Mpc, with an effective spherically-averaged  radius of 175 \h\ Mpc. Given the size of the MDPL2 simulation, it is possible to generate multiple independent (non-overlapping) finite volume realizations. For each sphere, we ignore any galaxy that exists outside of the sphere, and assign $\delta = 0$ for points outside the sphere. Hence each sphere represents a local universe realization: it can be over- or underdense, so we renormalize $\delta$ using the mean density within the sphere instead of the simulation box average.
For each sphere, we then calculate \betaf\ and $\sigma\sbr{8,g}$ using only that sphere's galaxies, although $\sigma_{8,m}$, which appears in the denominator of \bstar, continues to be calculated from the full simulation. As in real analyses, to account for the missing contribution from structures beyond the sphere's edge, in addition to fitting the \betaf, we also fit a residual bulk flow term.

Figure \ref{fig:slope_as_fn_sc} demonstrates how \bstar\ and $\sigma\sbr{8,g}$ depend on $R\sbr{max}$ for the case with no mass cut on the galaxy catalogue. As expected, as the size of the sphere increases, results converge to the values found for the full simulation for $R\sbr{max} \geq 200 \h$ Mpc.

There is a significant amount of scatter in the measured \sig\ values for small $R\sbr{max}$, varying by as much as 10\% of the mean value, and with a mean value that tends to be between 4-8 \% lower than that of the full simulation. The bias in the mean may be due to renormalizing the density fluctuation, $\delta = (\rho - \bar{\rho})/\bar{\rho}$ with the \emph{local} $\bar{\rho}$ measured on the scale of the sphere, $R\sbr{max}$, thus effectively filtering out the contribution to \sig\ from large-scale waves. The scatter arises from sample variance effects,  which is a combination of cosmic variance in the underlying DM structures and stochasticity in the SHMR. This  decreases as $R\sbr{max}$ increases. 

Likewise, the locally-measured value of \betaf\ also varies from sphere to sphere. Again, there are several reasons for this. First, the local $\bar{\rho}$ averaged over the sphere will differ from the true average. This is important because, as noted above, it leads to a renormalization of $\delta$. Second, there is also stochasticity in the SHMR which may affect the locally-measured \betaf.  Of these two, the first effect is the dominant one: we find that the total scatter in $\beta/f$ for $R\sbr{max} = 150 \h$ Mpc at $R_G = 4 \h$ Mpc is typically 7.0\% and is mostly due to cosmic variance in $\bar{\rho}$. The scatter in $\beta/f$ dominates the scatter in $\sigma\sbr{8,g}$ (2.5\%) when they are combined in \bstar.

Figure \ref{fig:sc_150} demonstrates how the cut samples ($R\sbr{max} = 150 \h$ Mpc) depend on the  Gaussian smoothing radius $R_G$. In particular, when compared to Figure \ref{fig:pvs_from_observables} we find that the average \bstar\ for these volume-limited realizations are unbiased between 3.75-4.75 \h\ Mpc. The value of \bstar\ at $R_G = 4 \h$ Mpc ranges between 0.98-1.01. More importantly, the standard deviation in \bstar\ from sphere to sphere is 0.077 at $R\sbr{max} = 150 \h\ $ Mpc, although this declines significantly to 0.032 at 200 \h\ Mpc and 0.020 at 300 \h\ Mpc. This suggests that sample variance effects are the dominant uncertainty in $f\sig$ for current data sets.

The 1D velocity scatter as a function of $R_G$ has a similar shape as found previously, but is 40-50 km/s larger than for the full simulation box. This increase in the scatter is primarily due to the degradation of the predicted velocities as one approaches $R\sbr{max}$. When we compare the 1D $\sigma_v$  for the subsample of galaxies located within the inner half of the sphere's volume with those in the outer half at an $R_G = 4 \h$ Mpc, we find scatters of $\sim$ 165 km/s and $\sim$ 210 km/s, respectively. This increase is attributed to the fact that the velocity tracers near the outer edge have poorer predicted peculiar velocities because of unaccounted-for structures outside of the survey limits.

\
\section{Summary and Discussion}\label{sec:discussion}

The key results of this paper are as follows:
\begin{itemize}
    \item The velocities of DM haloes are well predicted by linear theory from the true density field with a Gaussian smoothing $R_G = 4 \h$ Mpc with a velocity scatter of 154 km/s. This is in agreement with Appendix A of \cite{carrick2015cosmological}.
    \item This can be understood because, in Fourier space, a Gaussian filter with $R_G = 3$ -- 4 \h\ Mpc is a good match to the cross-correlation function of the density and velocity fields. 
    \item The accuracy and precision of the linear theory predictions do not depend on the mass of the velocity tracer; there is no "velocity bias", except for clusters with $M\sbr{h} > 10^{14} \h \Msol$.
    \item If DM haloes are used as tracers of the density field, and one calculates $\sigma\sbr{8,h}$ of the halo-mass-weighted density field, then $\beta\sbr{h} \sigma\sbr{8,h}$ is a good estimator of $\fsig$.
    \item If noise is added to the DM halo masses, then \bstar\ is biased high by only a percent, for a 0.1 dex noise level.
    \item When galaxy luminosity or stellar mass are used for the density field, the values of \bstar\ indicate that the method is unbiased to within 5\%, depending on the semi-analytic galaxy formation model.
    \item When the density field is restricted to a finite volume, there is additional uncertainty due to cosmic variance, at the level of 7\% for a 150 \h Mpc sphere.
\end{itemize}

The results for \bstar\ calculated using the same tracers for the velocity and density field, are summarised in Table \ref{tbl:beta_4hMpc}. Overall we find that the method has $\sim 5$\% systematic uncertainties. This can be improved with semi-analytic galaxy formation models that more carefully match the real SHMR and its scatter. There is also uncertainty due to finite volumes and cosmic variance. 

Previous work has neglected the additional scatter due to the coupled effects of stochasticity in the galaxy-mass relation and the cosmic variance effect of finite volumes. For example, the 2M++ catalogue \citep{LavHud11} has an effective volume of 175 \h\ Mpc. The SAGE model, weighted by stellar mass, shows that, for 150 \h\ Mpc spheres, the uncertainty on $\beta\sbr{g} \sigma_{8,g} = \fsig$ is 7.7\%, while the expected (interpolated) value for a survey of 2M++'s size is 5.2\%.  This is slightly higher than the uncertainty estimated by \cite{carrick2015cosmological}, \cite{BorHudLav20} and \cite{SaiColMag20} who adopted a 4\% sampling variance uncertainty, plus observational errors in $\beta$ due to uncertainties in peculiar velocity measurements, which are subdominant.
For precise quantification of the biases and systematic uncertainties in \fsig\ derived from a specific survey, e.g. 2M++, the best approach to minimizing the systematic errors will be to create mock catalogues that mimic the geometry and selection of that particular survey.

The cosmic variance uncertainty can be reduced in the future with deeper, all-sky redshift surveys. For example, a survey extending to a redshift of 0.2 (600 \h Mpc) would have an uncertainty of only 0.4\% in the mean mass density and hence 0.5 -- 0.6\% in the luminosity (or stellar mass) density.
In the North, the DESI Bright Galaxy Survey \citep{DESISurvey}, will observe 10 million nearby galaxies. In the South, 4MOST \citep{4MOST} has the capability to survey large volumes in the nearby Universe. The future looks bright.

\section*{Acknowledgements}


MJH acknowledges support from an NSERC Discovery grant.

The CosmoSim database used in this paper is a service by the Leibniz-Institute for Astrophysics Potsdam (AIP). The MultiDark database was developed in cooperation with the Spanish MultiDark Consolider Project CSD2009-00064. The authors gratefully acknowledge the Gauss Centre for Supercomputing e.V. (www.gauss-centre.eu) and the Partnership for Advanced Supercomputing in Europe (PRACE, www.prace-ri.eu) for funding the MultiDark simulation project by providing computing time on the GCS Supercomputer SuperMUC at Leibniz Supercomputing Centre (LRZ, www.lrz.de).

The Bolshoi simulations have been performed within the Bolshoi project of the University of California High-Performance Astro Computing Center (UC-HiPACC) and were run at the NASA Ames Research Center.

\section*{DATA AVAILABILITY STATEMENT}
The  data  underlying  this  article are publicly  available  from  the  COSMOSIM  database \url{https://www.cosmosim.org/}, with their respective publications cited in section \ref{sec:sim_data}.




\bibliographystyle{mnras}
\bibliography{PVs} 




%



\bsp	
\label{lastpage}
\end{document}